\documentclass[conference]{IEEEtran}
\IEEEoverridecommandlockouts
\usepackage{cite}
\usepackage{amsmath,amssymb,amsfonts}
\usepackage{algorithmic}
\usepackage{graphicx}
\usepackage{textcomp}
\usepackage{url}
\usepackage{hyperref}
\usepackage{subcaption}
\usepackage{xcolor}
\def\BibTeX{{\rm B\kern-.05em{\sc i\kern-.025em b}\kern-.08em
    T\kern-.1667em\lower.7ex\hbox{E}\kern-.125emX}}
\begin{document}

\title{Demo: Real-time Generative Multicasting with On-Device Intent-aware Semantic Decomposition}

\author{
\IEEEauthorblockN{
Xinkai Liu,
Mahdi Boloursaz Mashhadi,
Yi Ma,
Rahim Tafazolli
}
\IEEEauthorblockA{5GIC \& 6GIC, Institute for Communication Systems (ICS), University of Surrey, Guildford, United Kingdom\\
Emails: \{xinkai.liu, m.boloursazmashhadi, y.ma, r.tafazolli\}@surrey.ac.uk}
}

\maketitle

\begin{abstract}
We present a demonstration for generative multicasting with on-device, intent-aware semantic decomposition\footnote{See GitHub repository at: \url{https://github.com/Xinkai-Liu/Demo-Real-time-Intent-aware-Semantic-Multicast}}. At the transmitter, DNN-based segmentation extracts a semantic map from the source video, decomposing it into multiple sub-signal classes based on multi-user receiver intents. The transmitter broadcasts the semantic map to all users over shared wireless/network resources, thereby utilizing orthogonal resources only to transmit the sub-signal classes intended for each user. Users partially reconstruct and partially synthesize the signal by combining the received intended classes with non-intended classes locally synthesized by a generative model from the semantic map. We derive the rate-distortion/perception curves for reconstruction/synthesis with the generative model, to adaptively set compression rates for the semantic map and intended classes. Generative multicasting  significantly reduces the wireless/network resources required for existing/emerging multimedia multicasting applications. The system is real-time on a Google Coral Edge TPU with 4 TOPS (int8)\footnote{\url{https://coral.ai/products/dev-board}}. This is the first demonstration of generative multicasting representing a substantial advancement in on-device generative SemCom.



\end{abstract}


\section{Introduction}
\begin{figure}[t]
\centering
{\includegraphics[width=\linewidth]{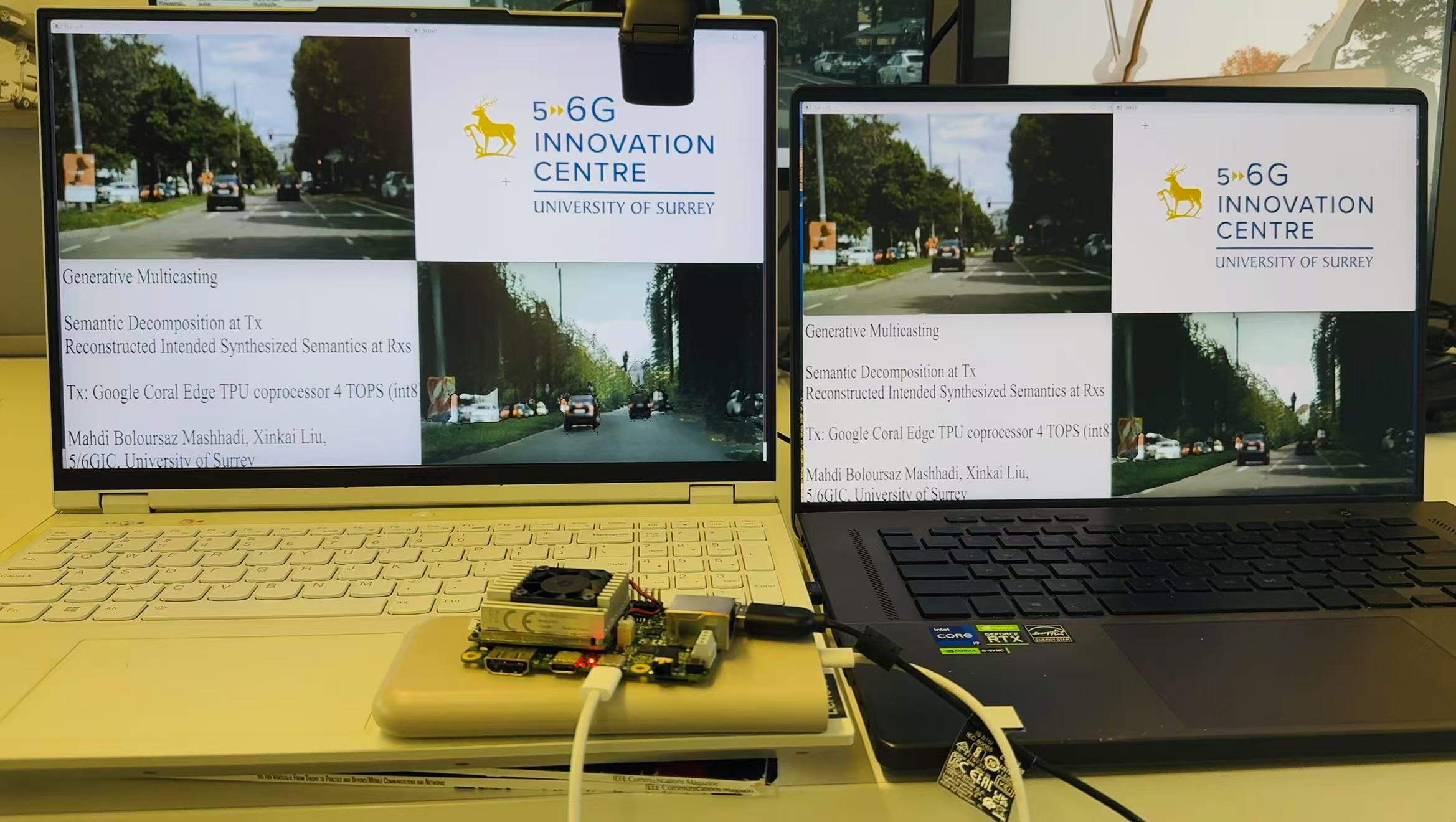}}
\vspace{-5mm}
\caption{\footnotesize Demonstration for real-time generative multicasting with
on-device intent-aware semantic decomposition. The transmitter is a Google Coral Edge TPU board with maximum 4 TOPS (int8) computations, equipped with a Debian Mendel Linux system, and powered by a portable power bank. The receivers are two laptops with NVIDIA RTX 3060 and RTX 4060 GPUs. The real-time video input is captured by a connected webcam, and all communications are through Wi-Fi 2x2 MIMO (802.11b/g/n/ac 2.4/5GHz). This is a ``live demo" with hardware/software set up at the conference venue.}
\vspace{-6mm}
\label{Demo_3}
\end{figure}

\textit{Semantic Communication (Semcom)} is poised crucial in next‐generation wireless networks, especially to reduce the data traffic for demanding new applications, e.g. the wireless metaverse, extended/mixed reality (XR/MR), and the internet of senses (IoS). In many such applications, multicasting of emerging multimedia signals is required, e.g. in XR/MR streaming the same content should be disseminated to multiple users simultaneously. More recently, \textit{Generative Semantic Communication (GenSC)} was proposed, where highly compressed semantics are extracted at the transmitter, communicated over the channel, and used at the receiver to guide a generative model to locally synthesize a semantically consistent and highly realistic signal, thereby achieving ultra-low-bitrate transmission \cite{Li2024generative, Xu2025generative}. Despite these developments, semantic communication in multiuser multicasting setups with user-specific intents and diverse user rate-distortion-perception \cite{RDP1, RDP2} requirements, remains less studied. Hence, we develop a prototype for \textit{generative semantic multicasting} with rate adaptation according to the multiuser intents and distortion/perception requirements. Our framework demonstrates low bitrate semantic multicasting, scalability to support many users, and compatibility with the existing wireless network designs, by a source–channel coding separation architecture with pre-training. Our proposed demo is displayed in Fig. \ref{Demo_3}.

\begin{figure*}[tb]
\centerline{\includegraphics[scale=0.25]{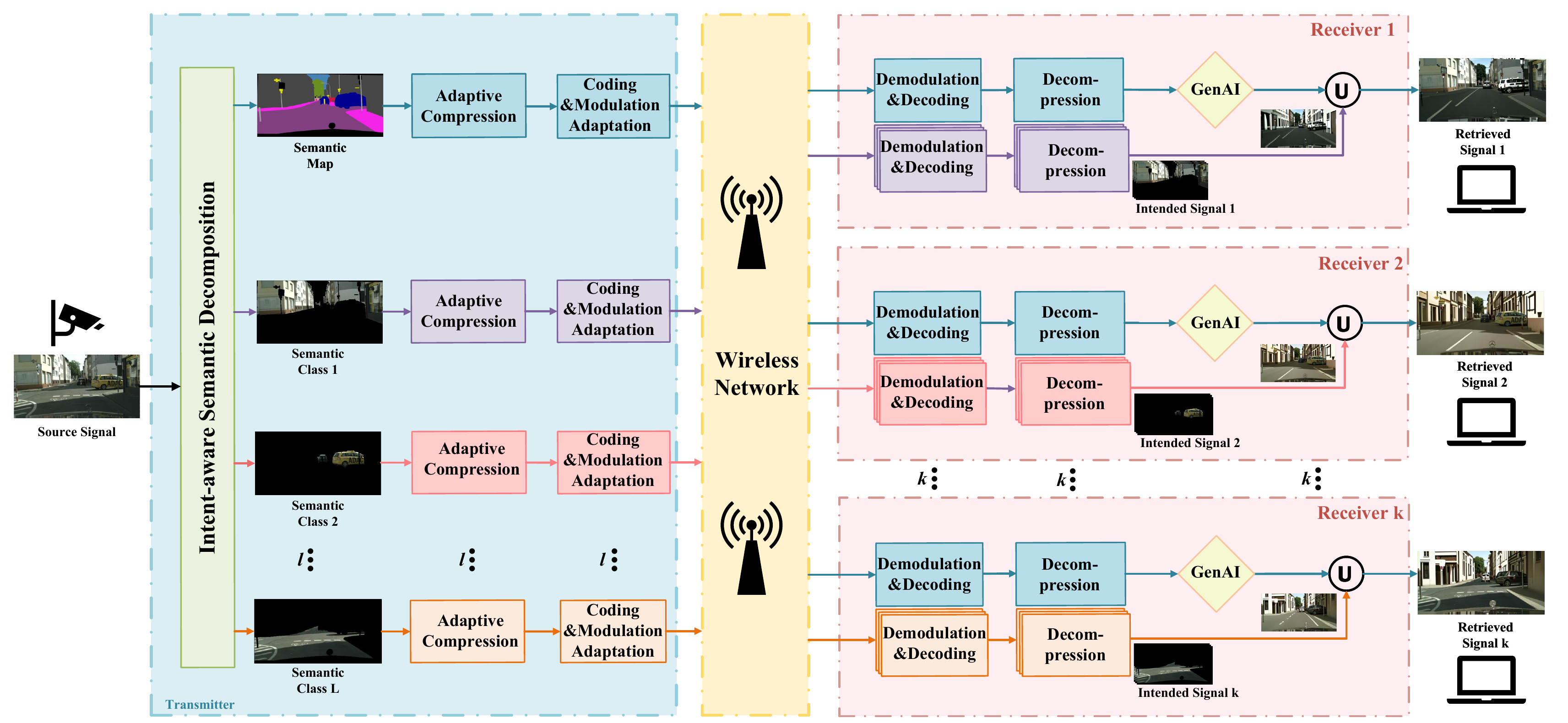}}
\vspace{-2mm}
\caption{\footnotesize Proposed Framework for Generative Semantic Multicasting with Intent-aware Semantic Decomposition.}
\vspace{-4mm}
\label{Demo_2}
\end{figure*}

\section{Technical Description}
\begin{figure}[!t]
\centering
{\includegraphics[scale=0.45]{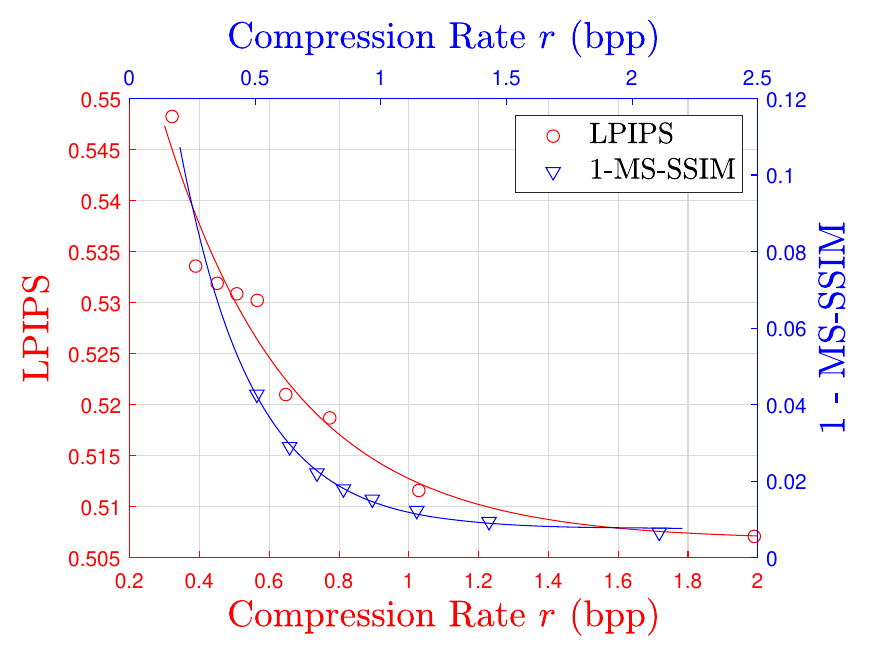}}
\vspace{-3mm}
\caption{\footnotesize Reconstruction/Synthesis Distortion/Perception curves.}
\vspace{-6mm}
\label{Demo_1}
\end{figure}

Fig. \ref{Demo_2} illustrates our proposed architecture for intent-aware generative semantic multicasting. The transmitter extracts a semantic map and decomposes the source signal into multiple sub-signal classes, leveraging the lightweight EdgeTPU-DeepLab-slim model\footnote{\url{https://github.com/tensorflow/models/blob/master/research/deeplab/g3doc/model_zoo.md\#edgetpu-deeplab-models-on-cityscapes}} optimized for real-time segmentation on the Edge TPU board. The transmitter then broadcasts the semantic map to all users over shared wireless/network resources, thereby utilizing orthogonal resources only to transmit the sub-signal classes intended for each user. For example, the sub-signal intended for a “traffic surveillance” user is the  “car” class but not “sky”. The “car” class will thereby be transmitted to this user over orthogonal channels to maintain reconstruction fidelity of the cars’ make, model, and number plate. This user will locally synthesize the other, i.e. non-intended, classes from the semantic map using a generative model. The multiuser intent is assumed available at the transmitter through feedback. To further reduce the data volume, JPEG compression with rate $r$ in (bpp), is applied to the semantic map and sub-signal classes intended for each user. For synthesis, we utilize the  GAN-based OASIS model\footnote{\url{https://github.com/boschresearch/OASIS}} with semantic guidance, that achieves high-fidelity, real-time performance. We pretrain the OASIS model on the Cityscapes dataset\footnote{\url{https://www.cityscapes-dataset.com/}}, and carry out extensive simulations to derive the approximate rate-distortion/perception curves for reconstruction $\Phi_r\left(r\right)=0.199 e^{(-3.454 r)} + 0.008$, and synthesis $\Phi_s\left(r\right)= 0.092 e^{(-2.732 r)} + 0.507$ via curve fitting in Fig. \ref{Demo_1}. The transmitter uses these curves to adaptively determine the compression rates for the semantic map and intended sub-signal classes, based on the multiuser requirements on the distortion/perception quality of the reconstructed/synthesized portion of the signals and the wireless/network conditions.

\begin{figure}[!ht]
  \centering
  \begin{subfigure}[t]{0.49\columnwidth}
    \centering
    \includegraphics[width=\linewidth]{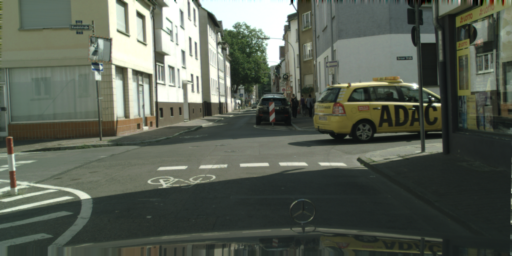}  
    \caption{\scriptsize Non-Generative Intent-unaware Multicasting, $R=814423$ and (MS-SSIM, LPIPS) $=(0.9977, - )$.}
    \label{visual1}
  \end{subfigure}
  \begin{subfigure}[t]{0.49\columnwidth}
    \centering
    \includegraphics[width=\linewidth]{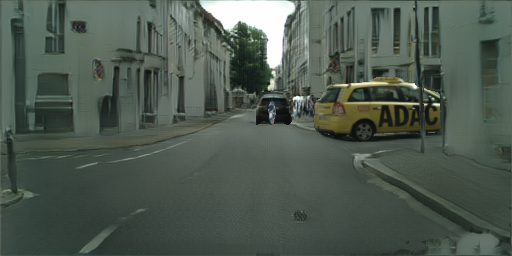}  
    \caption{\scriptsize Proposed framework with only the ``Car" class reconstructed, rest synthesized, $R=423703$, (MS-SSIM, LPIPS) $=(0.9977,0.5071)$.}
    \label{visual2}
  \end{subfigure}
  \begin{subfigure}[t]{0.49\columnwidth}
    \centering
    \includegraphics[width=\linewidth]{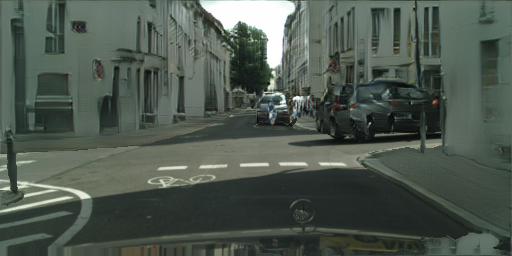}  
    \caption{\scriptsize Proposed framework with only the ``Road" class reconstructed, rest synthesized, $R=423703$, (MS-SSIM, LPIPS) $=(0.9977,0.5071)$.}
    \label{vi}
    \label{visual3}
  \end{subfigure}
  \begin{subfigure}[t]{0.49\columnwidth}
    \centering
    \includegraphics[width=\linewidth]{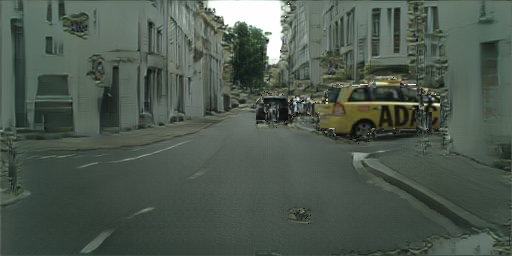}
    \caption{\scriptsize Rate-distortion-perception adaptation performance. The ``Car" class is reconstructed, rest synthesized, $R=88760$, (MS-SSIM, LPIPS) $=(0.9820,0.5309)$.}
    \label{vi}
    \label{visual4}
  \end{subfigure}
  \caption{\footnotesize Visual Quality Comparisons}
  \label{fig:visual_latency}
  \vspace{-7mm}
\end{figure}

\section{Results and Comparisons}
The total number of bits transmitted over the wireless network for 2 users is $R=423703$ with the proposed framework, while a conventional Non-Generative intent-unaware Multicasting (NGM) requires a total $R=814423$ bits for the same video at the same distortion level. Our proposed generative multicasting framework thereby significantly reduces the number of bits transmitted over the wireless network, specifically achieving a {48.0\%} reduction in comparison with NGM in the set up of this demo. Finally, in Fig. {\ref{fig:visual_latency}}, we demonstrate the visual quality of the retrieved signals for various distortion/perception levels, demonstrating the rate-distortion-perception adaptability of our proposed framework. Finally, note that although the demo is for 2 users due to hardware/space constraints for demonstration, but the proposed generative semantic multicasting framework is simply scalable to many users. A remaining challenge for large scale deployment is synchronization of the two data streams for reconstruction/synthesis at each user, which will be studied in our future work.
\vspace{-3mm}

\bibliographystyle{IEEEtran}
\bibliography{Demo}
\end{document}